\documentclass[showpacs,aps,graphicx,twocolumn]{revtex4}
\usepackage{graphicx}

\begin{document}

\title{Two-step complete polarization logic Bell-state analysis}

\author{Yu-Bo Sheng,$^{1}$\footnote{Email address:
shengyb@njupt.edu.cn} Lan Zhou$^{2}$}
\address{$^1$ Institute of Signal Processing  Transmission, Nanjing
University of Posts and Telecommunications, Nanjing, 210003,  China\\
 $^2$College of Mathematics \& Physics, Nanjing University of Posts and Telecommunications, Nanjing,
210003, China\\}

\date{\today }

\begin{abstract}
Logic qubit entanglement, which is also called the concatenated
Greenberger-Horne-Zeilinger (C-GHZ) state, is robust in practical noisy environment. In this paper, we will describe
an efficient approach to realize the complete polarization Bell-state analysis which is encoded in the logic qubit. We showed that the logic Bell-state
can be distinguished in two steps with the help of the parity-check measurement (PCM), which is constructed by the cross-Kerr nonlinearity.
We also explain that this approach can be used to distinguish arbitrary C-GHZ state with $N$ logic qubits. This protocol is useful in
the long-distance quantum communication based on the logic qubit entanglement.
\end{abstract}
\pacs{ 03.67.Dd, 03.67.Hk, 03.65.Ud} \maketitle

\section{Introduction}
Bell-state analysis is of vice importance in current quantum communication. Quantum teleportation \cite{teleportation}, quantum key distribution \cite{Ekert91},  quantum dense coding \cite{densecoding1},
quantum state sharing \cite{QSS1,QSS2}, and quantum secure direct communication \cite{QSDC1,QSDC2}  all need the Bell-state analysis.
Especially, in long-distance quantum communication, people should set up the long-distance quantum channel first. In a practical application,
they should exploit the entanglement swapping to extend the length of the entanglement, which is called the quantum repeaters \cite{repeater}. The key element of the quantum
repeaters is still the Bell-state analysis.

Usually, in an optical system, there are three different approaches to realize the Bell-state analysis which is encoded in the polarization degree of freedom.
The first approach requires the linear optical elements \cite{bell1,bell2,bell3,bell4,bell5,bell6}. However, it is showed that one cannot perform the complete Bell-state analysis with only linear optics. The optimal
success probability is only 50\% \cite{bell1,bell2,bell3}.  The second approach still requires the linear optical elements but resorts to the
hyperentanglement \cite{bell7,bell8,bell9,bell10,bell11,bell12}. For example, if we want to
distinguish the four Bell states encoded in the polarization degree of freedom. We first prepare the whole state in hyperentanglement, which entangles in both
polarization-spatial modes \cite{bell8}, or polarization-orbital-angular-momentum  degrees of freedom \cite{bell12}, simultaneously. By introducing other degree of freedom, the polarization Bell states can be completely discriminated.
In essence, this approach works in a large Hilbert space in two degrees of freedom. The complete Bell-state analysis with hyperentanglement has been well discussed and realized \cite{bell10,bell11,bell12}.
The third approach works in a nonlinear optical system \cite{bell13,bell14,bell15,bell16}. For instance, with the help of the cross-Kerr nonlinearity, they can construct the quantum nondemolition measurement,
such as the parity-check measurement (PCM) \cite{bell13}. The PCM can distinguish the even parity states $|H\rangle|H\rangle$ and $|V\rangle|V\rangle$ from the odd parity states
$|H\rangle|V\rangle$ and $|V\rangle|H\rangle$ deterministically. Here $|H\rangle$ is the horizonal polarized photon and  $|V\rangle$ is the vertical polarized photon, respectively.
In this way, the complete polarization Bell-state analysis can be well performed in two steps. The first step is to distinguish $|\phi^{\pm}\rangle$ from
$|\psi^{\pm}\rangle$. The second step is to distinguish $|\phi^{+}\rangle$ from $|\phi^{-}\rangle$, and $|\psi^{+}\rangle$ from $|\psi^{-}\rangle$, respectively. Here
\begin{eqnarray}
|\phi^{\pm}\rangle=\frac{1}{\sqrt{2}}(|H\rangle|H\rangle\pm|V\rangle|V\rangle),\nonumber\\
|\psi^{\pm}\rangle=\frac{1}{\sqrt{2}}(|H\rangle|V\rangle\pm|V\rangle|H\rangle).
\end{eqnarray}

On the other hand, it is known that the decoherence is one of the main obstacles in long-distance quantum communication. In the past decades, people developed
serval approaches to resist the decoherence. For example, they presented the quantum repeaters \cite{repeater,singlerepeater3} and nonlinear photon
amplification \cite{NLA1,NLA2,NLA3,NLA4} to resist the photon loss during the
photon distribution. They also proposed the entanglement purification \cite{C.H.Bennett1,Pan1,shengpra,shengpra2,wangc1,dengonestep1}
and concentration \cite{C.H.Bennett2,Yamamoto1,zhao1,shengsingle,shengWthree,dengsingle}to improve the quality of the degraded entanglement.
In current quantum communication protocols, they all encode the quantum qubit in the physical qubit directly, such as the polarization
time-bin, spatial modes degrees of freedom, and so on. Recently, Fr\"{o}wis and D\"{u}r developed a class of quantum entanglement, which is
encoded many physical qubit in a logic qubit \cite{cghz1}.  Such logic qubit entanglement has the similar feature
as the Greenberger-Horne-Zeiglinger (GHZ) state, but is more robust than the normal GHZ state in a noisy environment.
The logic qubit entanglement, which is also called the concatenated GHZ (C-GHZ) state can be described as \cite{cghz1,cghz2,cghz3,cghz4,cghz5,cghz6}
\begin{eqnarray}
|\Phi_{1}^{\pm}\rangle_{N,M}=\frac{1}{\sqrt{2}}(|GHZ^{+}_{M}\rangle^{\otimes N} \pm |GHZ^{-}_{M}\rangle^{\otimes N}).\label{logic}
\end{eqnarray}
Here $N$ is the number of logic qubit and $M$ is the number of the physical qubit in each logic qubit.
$|GHZ^{\pm}_{M}\rangle$ are the $M$-photon polarized GHZ states which can be written as
\begin{eqnarray}
|GHZ^{\pm}_{M}\rangle=\frac{1}{\sqrt{2}}(|H\rangle^{\otimes M}\pm|V\rangle^{\otimes M}).
\end{eqnarray}
 In 2014, Lu \emph{et al.}
realized the C-GHZ state with $N=2$ and $M=3$ in a linear optical system \cite{cghz6}. They also verified its robustness in a noisy environment.

As the logic entangled state is more robust than the entanglement which is encoded in the physical qubit directly, it is possible to perform the
quantum communication based on the logic qubit entanglement. In this paper, we will describe an approach to realize the complete Bell-state analysis and
GHZ analysis based on
the logic qubit entanglement. In our protocol, we exploit the cross-Kerr nonlinearity to construct the PCM gate to complete the task.
It is shown that the logic Bell-state analysis can be achieved in two steps.
We also show that this approach can be used to perform the arbitrary  C-GHZ state analysis.

This paper is organized as follows. In Sec. II, we first briefly introduce the PCM gate constructed by the cross-Kerr nonlinearity.
In Sec. III, we will describe the approach of  logic Bell-state analysis based on the PCM gate. In Sec. IV, we extend this protocol to
distinguish the arbitrary C-GHZ state. In Sec. V, we will make a discussion. In Sec. VI, we will provide a conclusion.

\section{Parity check measurement gate}
Cross-Kerr nonlinearity provides us a powerful tool to construct the quantum nondemonlition measurement, which has been widely
used in quantum information processing. There are many researches based on the cross-Kerr nonlinearity, including the construction
of the controlled-not (CNOT) gate \cite{QND1,lin1}, performing the Bell-state analysis \cite{bell13,bell4},
realizing the entanglement purification \cite{shengpra,shengpra2} and concentration \cite{shengsingle,shengWthree,dengsingle}, and so on \cite{he1,qi}.

\begin{figure}[!h]
\begin{center}
\includegraphics[width=7cm,angle=0]{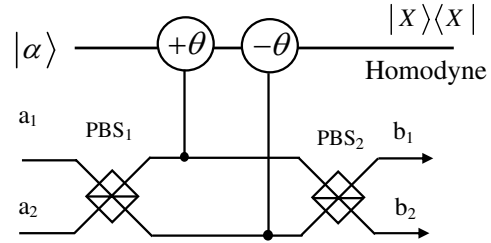}
\caption{A schematic drawing of our PCM gate. It can distinguish the even parity states $|H\rangle|H\rangle$ and
$|V\rangle|V\rangle$ from the odd parity states $|H\rangle|V\rangle$ and
$|V\rangle|H\rangle$. PBS represents the polarization beam splitters which can transmit the $|H\rangle$ photon and reflect the $|V\rangle$ photon.
Such PCM gate is also shown in Refs.\cite{shengpra,qi}.}
\end{center}
\end{figure}
In this section, we will briefly describe the PCM gate constructed by the cross-Kerr nonlinearity. As shown in Fig. 1, the
 Hamiltonian of a cross-Kerr nonlinear medium can be written as
$H=\hbar\chi \hat{n_{a}}\hat{n_{b}}$.  The $\hat{n_{a}}(\hat{n_{b}})$ is the number
operator for mode $a(b)$ \cite{QND1}.  The $\hbar\chi$ is the
coupling strength of the nonlinearity. It is decided by the
cross-Kerr material. If we consider a two-photon state
$|\varphi\rangle_{0}=\epsilon|H\rangle_{a_{1}}|H\rangle_{a_{2}}+\beta|H\rangle_{a_{1}}|V\rangle_{a_{2}}
+\gamma|V\rangle_{a_{1}}|H\rangle_{a_{2}}+\delta|V\rangle_{a_{1}}|V\rangle_{a_{2}}$.
Here $|\epsilon|^{2}+|\beta|^{2}+|\gamma|^{2}+|\delta|^{2}=1$ and $a_{1}(a_{2})$ is the spatial mode as shown in Fig.1.
The $|\varphi\rangle_{0}$ combined with the coherent state $|\alpha\rangle$
can be described as
\begin{eqnarray}
&&|\varphi\rangle_{0}|\alpha\rangle=(\epsilon|H\rangle_{a_{1}}|H\rangle_{a_{2}}+\beta|H\rangle_{a_{1}}|V\rangle_{a_{2}}\nonumber\\
&+&\gamma|V\rangle_{a_{1}}|H\rangle_{a_{2}}+\delta|V\rangle_{a_{1}}|V\rangle_{a_{2}})|\alpha\rangle\nonumber\\
&\rightarrow&(\epsilon|H\rangle_{b_{1}}|H\rangle_{b_{2}}+\delta|V\rangle_{b_{1}}|V\rangle_{b_{2}})|\alpha\rangle\nonumber\\
&+&\beta|H\rangle_{b_{1}}|V\rangle_{b_{2}}|\alpha e^{-i2\theta}\rangle+\gamma|V\rangle_{b_{1}}|H\rangle_{b_{2}}|\alpha e^{i2\theta}\rangle.\label{coherent}
\end{eqnarray}
The PCM gate works as follows. From Eq. (\ref{coherent}), if the coherent state picks up no phase shift, the state will become the even parity state
$\epsilon|H\rangle_{b_{1}}|H\rangle_{b_{2}}+\delta|V\rangle_{b_{1}}|V\rangle_{b_{2}}$. If the coherent state picks up the phase shift $2\theta$, the state will
collapse to the odd parity state $\beta|H\rangle_{a_{1}}|V\rangle_{a_{2}}
+\gamma|V\rangle_{a_{1}}|H\rangle_{a_{2}}$. Here we should require the $\pm2\theta$ undistinguished, which can be completed by X quadrature
measurement. It can
be achieved by choosing the local oscillator phase $\pi/2$
offset from the probe phase \cite{QND1}.

\section{Logic Bell-state analysis}
In this section, we will start to explain our logic Bell-state analysis.
The logic Bell state can be regarded as the special state of the  C-GHZ state with $N=M=2$ in Eq. (\ref{logic}). The logic Bell state contains two logic qubits.
Each logic qubit is encoded in  a polarized Bell states.
The four logic Bell states can be described as
\begin{eqnarray}
|\Phi^{\pm}\rangle_{AB}=\frac{1}{\sqrt{2}}(|\phi^{+}\rangle_{A}|\phi^{+}\rangle_{B}\pm|\phi^{-}\rangle_{A}|\phi^{-}\rangle_{B}),\nonumber\\
|\Psi^{\pm}\rangle_{AB}=\frac{1}{\sqrt{2}}(|\phi^{+}\rangle_{A}|\phi^{-}\rangle_{B}\pm|\phi^{-}\rangle_{A}|\phi^{+}\rangle_{B}).\label{bell1}
\end{eqnarray}

\begin{figure}[!h]
\begin{center}
\includegraphics[width=7cm,angle=0]{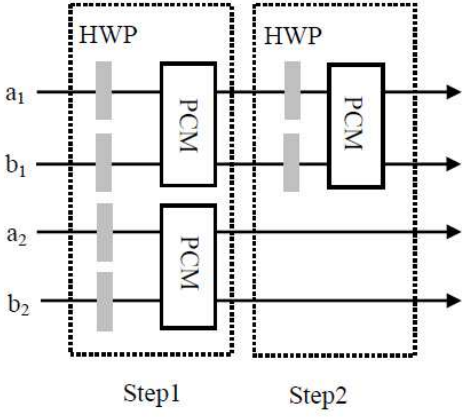}
\caption{A schematic drawing of our logic Bell-state analysis. PCM represents  the parity-check measurement gate described in Fig. 1. }
\end{center}
\end{figure}
From Fig. 2, the two photons in logic qubit A are in the spatial modes a$_{1}$ and a$_{2}$, respectively, and the two photons
in logic qubit B  are in the spatial modes b$_{1}$ and b$_{2}$, respectively. We first let four photons pass through the half wave plate (HWP),
which will make $|H\rangle\rightarrow\frac{1}{\sqrt{2}}(|H\rangle+|V\rangle)$, and $|V\rangle\rightarrow\frac{1}{\sqrt{2}}(|H\rangle-|V\rangle)$. The HWPs act
as the role of Hadamard operation.
The four HWPs will transform the states in Eq. (\ref{bell1}) to
\begin{eqnarray}
|\Phi^{\pm}\rangle_{AB}=\frac{1}{\sqrt{2}}(|\phi^{+}\rangle_{A}|\phi^{+}\rangle_{B}\pm|\psi^{+}\rangle_{A}|\psi^{+}\rangle_{B}),\nonumber\\
|\Psi^{\pm}\rangle_{AB}=\frac{1}{\sqrt{2}}(|\phi^{+}\rangle_{A}|\psi^{+}\rangle_{B}\pm|\psi^{+}\rangle_{A}|\phi^{+}\rangle_{B}).\label{bell}
\end{eqnarray}
After passing through the HWPs, the state $|\Phi^{+}\rangle_{AB}$ can be described as
\begin{eqnarray}
&&|\Phi^{+}\rangle_{AB}=\frac{1}{\sqrt{2}}(|\phi^{+}\rangle_{A}|\phi^{+}\rangle_{B}\pm|\psi^{+}\rangle_{A}|\psi^{+}\rangle_{B})\nonumber\\
&=&\frac{1}{\sqrt{2}}[\frac{1}{\sqrt{2}}(|H\rangle_{a_{1}}|H\rangle_{a_{2}}+|V\rangle_{a_{1}}|V\rangle_{a_{2}})\nonumber\\
&\otimes&\frac{1}{\sqrt{2}}(|H\rangle_{b_{1}}|H\rangle_{b_{2}}+|V\rangle_{b_{1}}|V\rangle_{b_{2}})\nonumber\\
&+&\frac{1}{\sqrt{2}}(|H\rangle_{a_{1}}|V\rangle_{a_{2}}+|V\rangle_{a_{1}}|H\rangle_{a_{2}})\nonumber\\
&\otimes&\frac{1}{\sqrt{2}}(|H\rangle_{b_{1}}|V\rangle_{a_{2}}+|V\rangle_{a_{1}}|H\rangle_{a_{2}}]\nonumber\\
&=&\frac{1}{2\sqrt{2}}[(|H\rangle_{a_{1}}|H\rangle_{a_{2}}|H\rangle_{b_{1}}|H\rangle_{b_{2}}
+|H\rangle_{a_{1}}|H\rangle_{a_{2}}|V\rangle_{b_{1}}|V\rangle_{b_{2}}\nonumber\\
&+&|V\rangle_{a_{1}}|V\rangle_{a_{2}}|H\rangle_{b_{1}}|H\rangle_{b_{2}}
+|V\rangle_{a_{1}}|V\rangle_{a_{2}}|V\rangle_{b_{1}}|V\rangle_{b_{2}})\nonumber\\
&+&|H\rangle_{a_{1}}|V\rangle_{a_{2}}|H\rangle_{b_{1}}|V\rangle_{b_{2}}+|H\rangle_{a_{1}}|V\rangle_{a_{2}}|V\rangle_{b_{1}}|H\rangle_{b_{2}}\nonumber\\
&+&|V\rangle_{a_{1}}|H\rangle_{a_{2}}|H\rangle_{b_{1}}|V\rangle_{b_{2}}+|V\rangle_{a_{1}}|H\rangle_{a_{2}}|V\rangle_{b_{1}}|H\rangle_{b_{2}}].\label{1}
\end{eqnarray}
In the first step, we let the four photons pass through the two PCM gates, respectively. Interestingly, the results of the two PCMs are the same.
They are both in the even parity or the odd parity. If both the PCM results are even, Eq. (\ref{1}) will become
\begin{eqnarray}
&&\rightarrow|H\rangle_{a_{1}}|H\rangle_{a_{2}}|H\rangle_{b_{1}}|H\rangle_{b_{2}}+|V\rangle_{a_{1}}|V\rangle_{a_{2}}|V\rangle_{b_{1}}|V\rangle_{b_{2}}\nonumber\\
&&+|H\rangle_{a_{1}}|V\rangle_{a_{2}}|H\rangle_{b_{1}}|V\rangle_{b_{2}}+|V\rangle_{a_{1}}|H\rangle_{a2}|V\rangle_{b_{1}}|H\rangle_{b_{2}}\nonumber\\
&&=|\phi^{+}\rangle_{a_{1}b_{1}}|\phi^{+}\rangle_{a_{2}b_{2}}.\label{collapse1}
\end{eqnarray}
On the other hand, if both the PCM results are odd, they will obtain
\begin{eqnarray}
&&\rightarrow|H\rangle_{a_{1}}|H\rangle_{a_{2}}|V\rangle_{b_{1}}|V\rangle_{b_{2}}+|V\rangle_{a_{1}}|V\rangle_{a_{2}}|H\rangle_{b_{1}}|H\rangle_{b_{2}}\nonumber\\
&&|H\rangle_{a_{1}}|V\rangle_{a_{2}}|V\rangle_{b_{1}}|H\rangle_{b_{2}}+|V\rangle_{a_{1}}|H\rangle_{a_{2}}|H\rangle_{b_{1}}|V\rangle_{b_{2}}\nonumber\\
&&=|\psi^{+}\rangle_{a_{1}b_{1}}|\psi^{+}\rangle_{a_{2}b_{_{2}}}.\label{collapse2}
\end{eqnarray}
Interestingly, if the initial state is $|\Phi^{-}\rangle_{AB}$, they can obtain the same results as $|\Phi^{+}\rangle_{AB}$. The PCM results are both
even or odd. In detail, if they are even, the $|\Phi^{-}\rangle_{AB}$ will collapse to
\begin{eqnarray}
&&\rightarrow|H\rangle_{a_{1}}|H\rangle_{a_{2}}|H\rangle_{b_{1}}|H\rangle_{b_{2}}+|V\rangle_{a_{1}}|V\rangle_{a_{2}}|V\rangle_{b_{1}}|V\rangle_{b_{2}}\nonumber\\
&&-|H\rangle_{a_{1}}|V\rangle_{a_{2}}|H\rangle_{b_{1}}|V\rangle_{b_{2}}-|V\rangle_{a_{1}}|H\rangle_{a_{2}}|V\rangle_{b_{1}}|H\rangle_{b_{2}}\nonumber\\
&&=|\phi^{-}\rangle_{a_{1}b_{1}}|\phi^{-}\rangle_{a_{2}b_{2}}.\label{collapse3}
\end{eqnarray}
On the other hand, if the measurement results are both odd, they will obtain
\begin{eqnarray}
&&\rightarrow|H\rangle_{a_{1}}|H\rangle_{a_{2}}|V\rangle_{b_{1}}|V\rangle_{b_{2}}+|V\rangle_{a_{1}}|V\rangle_{a_{2}}|H\rangle_{b_{1}}|H\rangle_{b_{2}}\nonumber\\
&&-|H\rangle_{a_{1}}|V\rangle_{a_{2}}|V\rangle_{b_{1}}|H\rangle_{b_{2}}-|V\rangle_{a_{1}}|H\rangle_{a_{2}}|H\rangle_{b_{1}}|V\rangle_{b_{2}}\nonumber\\
&&=|\psi^{-}\rangle_{a_{1}b_{1}}|\psi^{-}\rangle_{a_{2}b_{2}}.\label{collapse4}
\end{eqnarray}

If the initial state is $|\Psi^{+}\rangle_{AB}$ or, $|\Psi^{-}\rangle_{AB}$, after performing the PCM operations, the measurement results
are different. If the PCM in spatial modes a$_{1}$ and b$_{1}$ is even, the PCM in spatial modes a$_{1}$ and b$_{1}$ must be odd. On the other hand,
if the PCM in spatial modes a$_{1}$ and b$_{1}$ is odd, the PCM in spatial modes a$_{1}$ and b$_{1}$ must be even.
In the first case, the  $|\Psi^{+}\rangle_{AB}$ will collapse to $|\phi^{+}\rangle_{a_{1}b_{1}}|\psi^{+}\rangle_{a_{2}b_{2}}$ and $|\Psi^{-}\rangle_{AB}$
will collapse to $|\phi^{-}\rangle_{a_{1}b_{1}}|\psi^{-}\rangle_{a_{2}b_{2}}$. In the second case,   $|\Psi^{+}\rangle_{AB}$
will collapse to $|\psi^{+}\rangle_{a_{1}b_{1}}|\phi^{+}\rangle_{a_{2}b_{2}}$ and $|\Psi^{-}\rangle_{AB}$
will collapse to $|\psi^{-}\rangle_{a_{1}b_{1}}|\phi^{-}\rangle_{a_{2}b_{2}}$.

From above description, it is shown that the four logic Bell states can be divided into two groups according to the PCM results.
If two PCM results are the same, they are $|\Phi^{\pm}\rangle_{AB}$. If two PCM results are different, they are $|\Psi^{\pm}\rangle_{AB}$.
The next step is to distinguish   $|\Phi^{\pm}\rangle_{AB}$ or  $|\Psi^{\pm}\rangle_{AB}$ in each group. We take $|\Phi^{\pm}\rangle_{AB}$ for example.
From Eq. (\ref{collapse1}) and (\ref{collapse3}), if the PCM results are both even. The state in a$_{1}$b$_{1}$ must be
$|\phi^{+}\rangle_{a_{1}b_{1}}$,
if the initial state is $|\Phi^{+}\rangle_{AB}$. Otherwise, the state in a$_{1}$b$_{1}$ must be $|\phi^{-}\rangle_{a_{1}b_{1}}$,
if the initial state is $|\Phi^{-}\rangle_{AB}$. Therefore, the second step only need to distinguish the states $|\phi^{\pm}\rangle_{AB}$.
In the second step, after two photons passing
through the two HWPs, state $|\phi^{+}\rangle_{a_{1}b_{1}}$ does not change, while $|\phi^{-}\rangle_{a_{1}b_{1}}$ will become $|\psi^{+}\rangle_{a_{1}b_{1}}$.
Finally, by performing another PCM operation, if the PCM result is even, it must be $|\phi^{+}\rangle_{a_{1}b_{1}}$, and the initial state must be $|\Phi^{+}\rangle_{AB}$.
If the PCM result is odd, it must be $|\psi^{+}\rangle_{a_{1}b_{1}}$, and the initial state must be $|\Phi^{-}\rangle_{AB}$.
If the initial states are $|\Psi^{\pm}\rangle_{AB}$, they can be distinguished in the same way. In this way, the four logic Bell states can be completely distinguished.

\section{logic GHZ-state analysis}
It is straightforward to extend the approach of logic Bell-state analysis to the case of C-GHZ state.
Here we let the logic qubits are $|\phi^{+}\rangle$ and $|\phi^{-}\rangle$, respectively. Therefore, the arbitrary C-GHZ state can be described as
\begin{eqnarray}
&&|\Phi^{\pm}_{1}\rangle_{N,2}=\frac{1}{\sqrt{2}}(|\phi^{+}\rangle^{\otimes N}\pm|\phi^{-}\rangle^{\otimes N}),\nonumber\\
&&|\Phi^{\pm}_{2}\rangle_{N,2}=\frac{1}{\sqrt{2}}(|\phi^{-}\rangle|\phi^{+}\rangle^{\otimes N-1}\pm|\phi^{+}\rangle|\phi^{-}\rangle^{\otimes N-1}),\nonumber\\
&&\cdots\nonumber\\
&&|\Phi^{\pm}_{2^{N-1}}\rangle_{N,2}=\frac{1}{\sqrt{2}}(|\phi^{+}\rangle^{\otimes N-1}|\phi^{-}\rangle\pm|\phi^{-}\rangle^{\otimes N-1}|\phi^{+}\rangle.\nonumber\\\label{multi2}
\end{eqnarray}

\begin{figure}[!h]
\begin{center}
\includegraphics[width=9cm,angle=0]{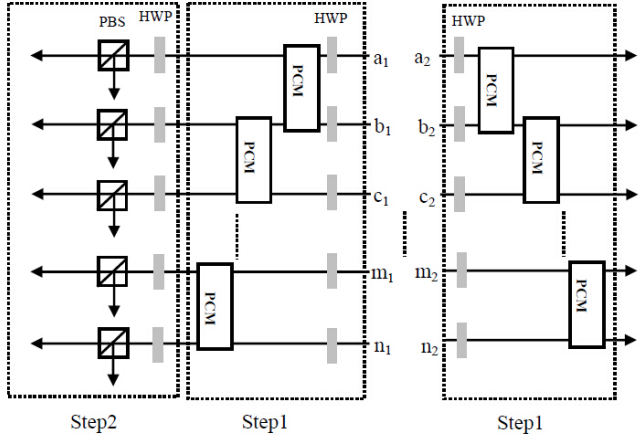}
\caption{A schematic drawing of distinguishing the C-GHZ state. }
\end{center}
\end{figure}

From Fig. 3, we first perform the Hadamard operation on each photon and make the states in Eq. (\ref{multi2}) become
\begin{eqnarray}
&&|\Phi^{\pm}_{1}\rangle_{N,2}=\frac{1}{\sqrt{2}}(|\phi^{+}\rangle^{\otimes N}\pm|\psi^{+}\rangle^{\otimes N}),\nonumber\\
&&|\Phi^{\pm}_{2}\rangle_{N,2}=\frac{1}{\sqrt{2}}(|\psi^{+}\rangle|\phi^{+}\rangle^{\otimes N-1}\pm|\phi^{+}\rangle|\psi^{+}\rangle^{\otimes N-1}),\nonumber\\
&&\cdots\nonumber\\
&&|\Phi^{\pm}_{2^{N-1}}\rangle_{N,2}=\frac{1}{\sqrt{2}}(|\phi^{+}\rangle^{\otimes N-1}|\psi^{+}\rangle\pm|\psi^{+}\rangle^{\otimes N-1}|\phi^{+}\rangle.\nonumber\\\label{multi3}
\end{eqnarray}
In order to describe this protocol clearly, we first discuss a simple case with $N=3$. As shown in Fig. 3, if $N=3$,
the eight C-GHZ states can be described as
\begin{eqnarray}
&&|\Phi^{\pm}_{1}\rangle_{3,2}=\frac{1}{\sqrt{2}}(|\phi^{+}\rangle_{A}|\phi^{+}\rangle_{B}|\phi^{+}\rangle_{C}
\pm|\psi^{+}\rangle_{A}|\psi^{+}\rangle_{B}|\psi^{+}\rangle_{C}),\nonumber\\
&&|\Phi^{\pm}_{2}\rangle_{3,2}=\frac{1}{\sqrt{2}}(|\psi^{+}\rangle_{A}|\phi^{+}\rangle_{B}|\phi^{+}\rangle_{C}
\pm|\phi^{+}\rangle_{A}|\psi^{+}\rangle_{B}|\psi^{+}\rangle_{C}),\nonumber\\
&&|\Phi^{\pm}_{3}\rangle_{3,2}=\frac{1}{\sqrt{2}}(|\phi^{+}\rangle_{A}|\psi^{+}\rangle_{B}|\phi^{+}\rangle_{C}
\pm|\psi^{+}\rangle_{A}|\phi^{+}\rangle_{B}|\psi^{+}\rangle_{C}),\nonumber\\
&&|\Phi^{\pm}_{4}\rangle_{3,2}=\frac{1}{\sqrt{2}}(|\phi^{+}\rangle_{A}|\phi^{+}\rangle_{B}|\psi^{+}\rangle_{C}
\pm|\psi^{+}\rangle_{A}|\psi^{+}\rangle_{B}|\phi^{+}\rangle_{C}).\nonumber\\
\end{eqnarray}

The state $|\Phi^{\pm}_{1}\rangle_{3,2}$ can be described as
\begin{eqnarray}
&&|\Phi^{\pm}_{1}\rangle_{3,2}=\frac{1}{\sqrt{2}}(|\phi^{+}\rangle_{A}|\phi^{+}\rangle_{B}|\phi^{+}\rangle_{C}
\pm|\psi^{+}\rangle_{A}|\psi^{+}\rangle_{B}|\psi^{+}\rangle_{C})\nonumber\\
&=&\frac{1}{\sqrt{2}}[\frac{1}{\sqrt{2}}(|H\rangle_{a_{1}}|H\rangle_{a_{2}}+|V\rangle_{a_{1}}|V\rangle_{a_{2}})\nonumber\\
&\otimes&\frac{1}{\sqrt{2}}(|H\rangle_{b_{1}}|H\rangle_{b_{2}}+|V\rangle_{b_{1}}|V\rangle_{b_{2}})\nonumber\\
&\otimes&\frac{1}{\sqrt{2}}(|H\rangle_{c_{1}}|H\rangle_{c_{2}}+|V\rangle_{c_{1}}|V\rangle_{c_{2}})\nonumber\\
&\pm&\frac{1}{\sqrt{2}}(|H\rangle_{a_{1}}|V\rangle_{a_{2}}+|V\rangle_{a_{1}}|H\rangle_{a_{2}})\nonumber\\
&\otimes&\frac{1}{\sqrt{2}}(|H\rangle_{b_{1}}|V\rangle_{a_{2}}+|V\rangle_{a_{1}}|H\rangle_{a_{2}}\nonumber\\
&\otimes&\frac{1}{\sqrt{2}}(|H\rangle_{c_{1}}|V\rangle_{c_{2}}+|V\rangle_{c_{1}}|H\rangle_{a_{2}}].
\end{eqnarray}
From Fig. 3, the basic principle of this protocol is to make the PCM operation between neighbor  physical qubit in each logic qubit.
In the case of $N=3$, we  first perform  the PCM operation between the photons in spatial modes $a_{1}b_{1}$, $b_{1}c_{1}$, $a_{2}b_{2}$, $b_{2}c_{2}$ respectively.
Interestingly, if the initial states are $|\Phi^{\pm}_{1}\rangle_{3,2}$, the result of PCM in the left side say  $a_{1}b_{1}$
is always the same as the result of the right side say $a_{2}b_{2}$.
The result of  PCM between $b_{1}c_{1}$ in the left side  is also the same as the result of $b_{2}c_{2}$ in the right side. We denote the even parity of the qubits in $a_{1}b_{1}$ modes
as $0_{a_{1}b_{1}}$ and the odd parity as $1_{a_{1}b_{1}}$, respectively. If the initial states are $|\Phi^{\pm}_{1}\rangle_{3,2}$,
all the possible cases of the PCM results can be written as $0_{a_{1}b_{1}}0_{b_{1}c_{1}}0_{a_{2}b_{2}}0_{b_{2}c_{2}}$,
$0_{a_{1}b_{1}}1_{b_{1}c_{1}}0_{a_{2}b_{2}}1_{b_{2}c_{2}}$,
  $1_{a_{1}b_{1}}0_{b_{1}c_{1}}1_{a_{2}b_{2}}0_{b_{2}c_{2}}$, or $1_{a_{1}b_{1}}1_{b_{1}c_{1}}1_{a_{2}b_{2}}1_{b_{2}c_{2}}$, respectively.
If the initial states are $|\Phi^{\pm}_{2}\rangle_{3,2}$, the PCM results can be written as $0_{a_{1}b_{1}}0_{b_{1}c_{1}}1_{a_{2}b_{2}}0_{b_{2}c_{2}}$,
 $0_{a_{1}b_{1}}1_{b_{1}c_{1}}1_{a_{2}b_{2}}1_{b_{2}c_{2}}$, $1_{a_{1}b_{1}}0_{b_{1}c_{1}}0_{a_{2}b_{2}}0_{b_{2}c_{2}}$,
 or $1_{a_{1}b_{1}}1_{b_{1}c_{1}}0_{a_{2}b_{2}}1_{b_{2}c_{2}}$. That is to say, the PCM in spatial modes $a_{1}b_{1}$ is always different from  $a_{2}b_{2}$, while the
PCM result in $b_{1}c_{1}$ is the same as that in  $b_{2}c_{2}$. If the initial state is $|\Phi^{\pm}_{3}\rangle_{3,2}$,
the PCM results in the left side are always different
from the right side. They are $0_{a_{1}b_{1}}0_{b_{1}c_{1}}1_{a_{2}b_{2}}1_{b_{2}c_{2}}$, $0_{a_{1}b_{1}}1_{b_{1}c_{1}}1_{a_{2}b_{2}}0_{b_{2}c_{2}}$,
 $1_{a_{1}b_{1}}0_{b_{1}c_{1}}0_{a_{2}b_{2}}1_{b_{2}c_{2}}$, or $1_{a_{1}b_{1}}1_{b_{1}c_{1}}0_{a_{2}b_{2}}0_{b_{2}c_{2}}$, respectively.
Finally, if the initial states are $|\Phi^{\pm}_{4}\rangle_{3,2}$, the PCM result of $a_{1}b_{1}$ is the same as $a_{2}b_{2}$, while the PCM of  $b_{1}c_{1}$
is different from $b_{2}c_{2}$. They are $0_{a_{1}b_{1}}0_{b_{1}c_{1}}0_{a_{2}b_{2}}1_{b_{2}c_{2}}$, $0_{a_{1}b_{1}}1_{b_{1}c_{1}}0_{a_{2}b_{2}}0_{b_{2}c_{2}}$,
$1_{a_{1}b_{1}}0_{b_{1}c_{1}}1_{a_{2}b_{2}}1_{b_{2}c_{2}}$, or $1_{a_{1}b_{1}}1_{b_{1}c_{1}}1_{a_{2}b_{2}}0_{b_{2}c_{2}}$, respectively.

From above description, we can find that the eight C-GHZ states can be divided into four groups, according to the PCM results. The four groups are
$\{|\Phi^{\pm}_{1}\rangle_{3,2}\}$, $\{|\Phi^{\pm}_{2}\rangle_{3,2}\}$, $\{|\Phi^{\pm}_{3}\rangle_{3,2}\}$ and $\{|\Phi^{\pm}_{4}\rangle_{3,2}\}$, respectively.
Therefore, the second step is to distinguish the two states in each group.
We first discuss $|\Phi^{\pm}_{1}\rangle_{3,2}$. If the initial states are $|\Phi^{\pm}_{1}\rangle_{3,2}$, the PCM results are one of
$0_{a_{1}b_{1}}0_{b_{1}c_{1}}0_{a_{2}b_{2}}0_{b_{2}c_{2}}$, $0_{a_{1}b_{1}}1_{b_{1}c_{1}}0_{a_{2}b_{2}}1_{b_{2}c_{2}}$,
$1_{a_{1}b_{1}}0_{b_{1}c_{1}}1_{a_{2}b_{2}}0_{b_{2}c_{2}}$, and $1_{a_{1}b_{1}}1_{b_{1}c_{1}}1_{a_{2}b_{2}}1_{b_{2}c_{2}}$,
with the equal probability of $\frac{1}{4}$, respectively.
First, if the PCM result is $0_{a_{1}b_{1}}0_{b_{1}c_{1}}0_{a_{2}b_{2}}0_{b_{2}c_{2}}$,
the states  $|\Phi^{\pm}_{1}\rangle_{3,2}$ will become
\begin{eqnarray}
&&|\Phi^{\pm}_{1}\rangle_{3,2}\rightarrow\frac{1}{2}[(|H\rangle_{a_{1}}|H\rangle_{a_{2}}|H\rangle_{b_{1}}|H\rangle_{b_{2}}|H\rangle_{c_{1}}|H\rangle_{c_{2}}\nonumber\\
&+&|V\rangle_{a_{1}}|V\rangle_{a_{2}}|V\rangle_{b_{1}}|V\rangle_{b_{2}}|V\rangle_{c_{1}}|V\rangle_{c_{2}})\nonumber\\
&\pm&(|H\rangle_{a_{1}}|V\rangle_{a_{2}}|H\rangle_{b_{1}}|V\rangle_{b_{2}}|H\rangle_{c_{1}}|V\rangle_{c_{2}}\nonumber\\
&+&|V\rangle_{a_{1}}|H\rangle_{a_{2}}|V\rangle_{b_{1}}|H\rangle_{b_{2}}|V\rangle_{c_{1}}|H\rangle_{c_{2}})]\nonumber\\
&=&\frac{1}{\sqrt{2}}(|H\rangle_{a_{1}}|H\rangle_{b_{1}}|H\rangle_{c_{1}}\pm|V\rangle_{a_{1}}|V\rangle_{b_{1}}|V\rangle_{c_{1}})\nonumber\\
&\otimes&\frac{1}{\sqrt{2}}(|H\rangle_{a_{2}}|H\rangle_{b_{2}}|H\rangle_{c_{2}}\pm|V\rangle_{a_{2}}|V\rangle_{b_{2}}|V\rangle_{c_{2}}).
\end{eqnarray}
Second, if the PCM is $0_{a_{1}b_{1}}1_{b_{1}c_{1}}0_{a_{2}b_{2}}1_{b_{2}c_{2}}$, it will make $|\Phi^{\pm}_{1}\rangle_{3,2}$ become
\begin{eqnarray}
&&|\Phi^{\pm}_{1}\rangle_{3,2}\rightarrow\frac{1}{\sqrt{2}}(|H\rangle_{a_{1}}|H\rangle_{b_{1}}|V\rangle_{c_{1}}\pm|V\rangle_{a_{1}}|V\rangle_{b_{1}}|H\rangle_{c_{1}})\nonumber\\
&\otimes&\frac{1}{\sqrt{2}}(|H\rangle_{a_{2}}|H\rangle_{b_{2}}|V\rangle_{c_{2}}\pm|V\rangle_{a_{2}}|V\rangle_{b_{2}}|H\rangle_{c_{2}}).
\end{eqnarray}
Third, if the PCM is $1_{a_{1}b_{1}}0_{b_{1}c_{1}}1_{a_{2}b_{2}}0_{b_{2}c_{2}}$, it will make $|\Phi^{\pm}_{1}\rangle_{3,2}$ become
\begin{eqnarray}
&&|\Phi^{\pm}_{1}\rangle_{3,2}\rightarrow\frac{1}{\sqrt{2}}(|V\rangle_{a_{1}}|H\rangle_{b_{1}}|H\rangle_{c_{1}}\pm|V\rangle_{a_{1}}|H\rangle_{b_{1}}|H\rangle_{c_{1}})\nonumber\\
&\otimes&\frac{1}{\sqrt{2}}(|V\rangle_{a_{2}}|H\rangle_{b_{2}}|H\rangle_{c_{2}}\pm|V\rangle_{a_{2}}|H\rangle_{b_{2}}|H\rangle_{c_{2}}).
\end{eqnarray}
Forth, if the PCM result is $1_{a_{1}b_{1}}1_{b_{1}c_{1}}1_{a_{2}b_{2}}1_{b_{2}c_{2}}$, it will make $|\Phi^{\pm}_{1}\rangle_{3,2}$ become
\begin{eqnarray}
&&|\Phi^{\pm}_{1}\rangle_{3,2}\rightarrow\frac{1}{\sqrt{2}}(|H\rangle_{a_{1}}|V\rangle_{b_{1}}|H\rangle_{c_{1}}\pm|H\rangle_{a_{1}}|V\rangle_{b_{1}}|H\rangle_{c_{1}})\nonumber\\
&\otimes&\frac{1}{\sqrt{2}}(|H\rangle_{a_{2}}|V\rangle_{b_{2}}|H\rangle_{c_{2}}\pm|H\rangle_{a_{2}}|V\rangle_{b_{2}}|H\rangle_{c_{2}}).
\end{eqnarray}

The next step is only  to distinguish the states $\frac{1}{\sqrt{2}}(|H\rangle_{a_{1}}|H\rangle_{b_{1}}|H\rangle_{c_{1}}\pm|V\rangle_{a_{1}}|V\rangle_{b_{1}}|V\rangle_{c_{1}})$.
Certainly, if we obtain the other states, such as  $\frac{1}{\sqrt{2}}(|H\rangle_{a_{1}}|H\rangle_{b_{1}}|V\rangle_{c_{1}}\pm|V\rangle_{a_{1}}|V\rangle_{b_{1}}|H\rangle_{c_{1}})$ in the second case,
we can perform a bit-flip operation on the $c_{1}$ photon and make them become $\frac{1}{\sqrt{2}}(|H\rangle_{a_{1}}|H\rangle_{b_{1}}|H\rangle_{c_{1}}\pm|V\rangle_{a_{1}}|V\rangle_{b_{1}}|V\rangle_{c_{1}})$.

The discrimination of the states $\frac{1}{\sqrt{2}}(|H\rangle_{a_{1}}|H\rangle_{b_{1}}|H\rangle_{c_{1}}\pm|V\rangle_{a_{1}}|V\rangle_{b_{1}}|V\rangle_{c_{1}})$
 can be described as follows. As shown in Fig. 3, we first perform the Hadamard operations on each photons and make
  $\frac{1}{\sqrt{2}}(|H\rangle_{a_{1}}|H\rangle_{b_{1}}|H\rangle_{c_{1}}\pm|V\rangle_{a_{1}}|V\rangle_{b_{1}}|V\rangle_{c_{1}})$ become
 \begin{eqnarray}
&& \frac{1}{\sqrt{2}}(|H\rangle_{a_{1}}|H\rangle_{b_{1}}|H\rangle_{c_{1}}+|V\rangle_{a_{1}}|V\rangle_{b_{1}}|V\rangle_{c_{1}})\nonumber\\
 &&\rightarrow\frac{1}{2}(|H\rangle_{a_{1}}|H\rangle_{b_{1}}|H\rangle_{c_{1}}+|H\rangle_{a_{1}}|V\rangle_{b_{1}}|V\rangle_{c_{1}}\nonumber\\
 &+&|V\rangle_{a_{1}}|H\rangle_{b_{1}}|V\rangle_{c_{1}}+|V\rangle_{a_{1}}|V\rangle_{b_{1}}|H\rangle_{c_{1}}),
\end{eqnarray}
 and
 \begin{eqnarray}
   &&\frac{1}{\sqrt{2}}(|H\rangle_{a_{1}}|H\rangle_{b_{1}}|H\rangle_{c_{1}}-|V\rangle_{a_{1}}|V\rangle_{b_{1}}|V\rangle_{c_{1}})\nonumber\\
 &&\rightarrow\frac{1}{2}(|H\rangle_{a_{1}}|H\rangle_{b_{1}}|V\rangle_{c_{1}}+|H\rangle_{a_{1}}|V\rangle_{b_{1}}|H\rangle_{c_{1}}\nonumber\\
 &+&|V\rangle_{a_{1}}|H\rangle_{b_{1}}|H\rangle_{c_{1}}+|V\rangle_{a_{1}}|V\rangle_{b_{1}}|V\rangle_{c_{1}}).
\end{eqnarray}
Subsequently,  we let three photons pass through the polarization beam splitters (PBSs), respectively. The PBS will transmit the $|H\rangle$ polarized photon and
reflect the $|V\rangle$ polarized photon. Finally, by detecting the photons in each output modes, we can distinguish the state $\frac{1}{\sqrt{2}}(|H\rangle_{a_{1}}|H\rangle_{b_{1}}|H\rangle_{c_{1}}+|V\rangle_{a_{1}}|V\rangle_{b_{1}}|V\rangle_{c_{1}})$ from $\frac{1}{\sqrt{2}}(|H\rangle_{a_{1}}|H\rangle_{b_{1}}|H\rangle_{c_{1}}-|V\rangle_{a_{1}}|V\rangle_{b_{1}}|V\rangle_{c_{1}})$. If the number of $|V\rangle$ is even, it is $\frac{1}{\sqrt{2}}(|H\rangle_{a_{1}}|H\rangle_{b_{1}}|H\rangle_{c_{1}}+|V\rangle_{a_{1}}|V\rangle_{b_{1}}|V\rangle_{c_{1}})$, and the initial state
is $|\Phi^{+}_{1}\rangle_{3,2}$. Otherwise, if the number of $|V\rangle$ is odd, it is $\frac{1}{\sqrt{2}}(|H\rangle_{a_{1}}|H\rangle_{b_{1}}|H\rangle_{c_{1}}-|V\rangle_{a_{1}}|V\rangle_{b_{1}}|V\rangle_{c_{1}})$, and the initial state
is $|\Phi^{-}_{1}\rangle_{3,2}$.

So far, we have completely distinguished the states $|\Phi^{\pm}_{1}\rangle_{3,2}$. The other six states can be distinguished with the same principle.
For example, in the first step, if the PCM results are  $0_{a_{1}b_{1}}0_{b_{1}c_{1}}1_{a_{2}b_{2}}0_{b_{2}c_{2}}$, $0_{a_{1}b_{1}}1_{b_{1}c_{1}}1_{a_{2}b_{2}}1_{b_{2}c_{2}}$, $1_{a_{1}b_{1}}0_{b_{1}c_{1}}0_{a_{2}b_{2}}0_{b_{2}c_{2}}$, or $1_{a_{1}b_{1}}1_{b_{1}c_{1}}0_{a_{2}b_{2}}1_{b_{2}c_{2}}$, the initial states must be one of the states $|\Phi^{\pm}_{2}\rangle_{3,2}$. The second step
is to distinguish $|\Phi^{+}_{2}\rangle_{3,2}$ from $|\Phi^{-}_{2}\rangle_{3,2}$. We take the PCM result  $0_{a_{1}b_{1}}0_{b_{1}c_{1}}1_{a_{2}b_{2}}0_{b_{2}c_{2}}$ as an example. The other cases can be discussed with the same principle.
If the PCM result is $0_{a_{1}b_{1}}0_{b_{1}c_{1}}1_{a_{2}b_{2}}0_{b_{2}c_{2}}$, $|\Phi^{\pm}_{2}\rangle_{3,2}$ becomes
 \begin{eqnarray}
 &&|\Phi^{\pm}_{2}\rangle_{3,2}\rightarrow\frac{1}{2}[(|H\rangle_{a_{1}}|V\rangle_{a_{2}}|H\rangle_{b_{1}}|H\rangle_{b_{2}}|H\rangle_{c_{1}}|H\rangle_{c_{2}}\nonumber\\
&+&|V\rangle_{a_{1}}|H\rangle_{a_{2}}|V\rangle_{b_{1}}|V\rangle_{b_{2}}|V\rangle_{c_{1}}|V\rangle_{c_{2}})\nonumber\\
&\pm&(|H\rangle_{a_{1}}|H\rangle_{a_{2}}|H\rangle_{b_{1}}|V\rangle_{b_{2}}|H\rangle_{c_{1}}|V\rangle_{c_{2}}\nonumber\\
&+&|V\rangle_{a_{1}}|V\rangle_{a_{2}}|V\rangle_{b_{1}}|H\rangle_{b_{2}}|V\rangle_{c_{1}}|H\rangle_{c_{2}})]\nonumber\\
&=&\frac{1}{\sqrt{2}}(|H\rangle_{a_{1}}|H\rangle_{b_{1}}|H\rangle_{c_{1}}\pm|V\rangle_{a_{1}}|V\rangle_{b_{1}}|V\rangle_{c_{1}})\nonumber\\
&\otimes&\frac{1}{\sqrt{2}}(|H\rangle_{a_{2}}|V\rangle_{b_{2}}|V\rangle_{c_{2}}\pm|V\rangle_{a_{2}}|H\rangle_{b_{2}}|H\rangle_{c_{2}}).
\end{eqnarray}

 Similarly, the second step is also to distinguish the states
 $\frac{1}{\sqrt{2}}(|H\rangle_{a_{1}}|H\rangle_{b_{1}}|H\rangle_{c_{1}}\pm|V\rangle_{a_{1}}|V\rangle_{b_{1}}|V\rangle_{c_{1}})$.
Certainly, if the PCM result in the first step is  $0_{a_{1}b_{1}}1_{b_{1}c_{1}}1_{a_{2}b_{2}}1_{b_{2}c_{2}}$,
$1_{a_{1}b_{1}}0_{b_{1}c_{1}}0_{a_{2}b_{2}}0_{b_{2}c_{2}}$,
 or $1_{a_{1}b_{1}}1_{b_{1}c_{1}}0_{a_{2}b_{2}}1_{b_{2}c_{2}}$, it can also be distinguished
 with the same principle. In this way, we can be simplified to distinguish
  $\frac{1}{\sqrt{2}}(|H\rangle_{a_{1}}|H\rangle_{b_{1}}|H\rangle_{c_{1}}\pm|V\rangle_{a_{1}}|V\rangle_{b_{1}}|V\rangle_{c_{1}})$
  after performing a bit-flip operation in the next step.  Therefore, we can completely
 distinguish the states  $|\Phi^{\pm}_{1}\rangle_{3,2}$.  The other states   $|\Phi^{\pm}_{2}\rangle_{3,2}$ and $|\Phi^{\pm}_{3}\rangle_{3,2}$ can also be
 distinguished in the same way. If the initial states are  $|\Phi^{\pm}_{3}\rangle_{3,2}$, the PCM results in the first step must be
  $0_{a_{1}b_{1}}0_{b_{1}c_{1}}1_{a_{2}b_{2}}1_{b_{2}c_{2}}$, $0_{a_{1}b_{1}}1_{b_{1}c_{1}}1_{a_{2}b_{2}}0_{b_{2}c_{2}}$,
   $1_{a_{1}b_{1}}0_{b_{1}c_{1}}0_{a_{2}b_{2}}1_{b_{2}c_{2}}$,
   or $1_{a_{1}b_{1}}1_{b_{1}c_{1}}0_{a_{2}b_{2}}0_{b_{2}c_{2}}$, respectively. If the initial states are
 $|\Phi^{\pm}_{4}\rangle_{3,2}$, the PCM results in the first step must be $0_{a_{1}b_{1}}0_{b_{1}c_{1}}0_{a_{2}b_{2}}1_{b_{2}c_{2}}$,
  $0_{a_{1}b_{1}}1_{b_{1}c_{1}}0_{a_{2}b_{2}}0_{b_{2}c_{2}}$, $1_{a_{1}b_{1}}0_{b_{1}c_{1}}1_{a_{2}b_{2}}1_{b_{2}c_{2}}$, or $1_{a_{1}b_{1}}1_{b_{1}c_{1}}1_{a_{2}b_{2}}0_{b_{2}c_{2}}$, respectively.
Therefore, in the second step, we only need to distinguish the states
 $\frac{1}{\sqrt{2}}(|H\rangle_{a_{1}}|H\rangle_{b_{1}}|H\rangle_{c_{1}}\pm|V\rangle_{a_{1}}|V\rangle_{b_{1}}|V\rangle_{c_{1}})$ in each group.
 In this way, all eight states  $|\Phi^{\pm}_{1}\rangle_{3,2}$,  $|\Phi^{\pm}_{2}\rangle_{3,2}$,  $|\Phi^{\pm}_{3}\rangle_{3,2}$ and
 $|\Phi^{\pm}_{4}\rangle_{3,2}$ can be completely distinguished.

 It is straightforward to extend this protocol to distinguish the C-GHZ state with $N$ logic qubits as shown in Eq. (\ref{multi2}) or Eq. (\ref{multi3}).
 The basic principle is also shown in Fig. 3. In the first step, we perform the PCM operation on the photons in $a_{1}b_{1}$, $b_{1}c_{1}$, $\cdots$,
 $m_{1}n_{1}$ in the left side and $a_{2}b_{2}$, $b_{2}c_{2}$, $\cdots$,
 $m_{2}n_{2}$ in the right side. In each side, we should perform $N-1$ PCM operations. Interestingly, if the initial states are $|\Phi^{\pm}_{1}\rangle_{N,2}$,
 the PCM result in the left side always equals to the result in the right side in the correspond position. That is the PCM result in $a_{1}b_{1}$ equals to that in
 $a_{2}b_{2}$. The PCM result in $b_{1}c_{1}$ equals to that in $b_{2}c_{2}$, $\cdots$, and the PCM result in $m_{1}n_{1}$ equals to that in $m_{2}n_{2}$.
 The PCM results in the left side say $P_{a_{1}b_{1}}P_{b_{1}c_{1}}\cdots P_{m_{1}n_{1}} (P=0,1)$ have $2^{N-1}$ possible cases with the same probability $\frac{1}{2^{N-1}}$. They are
 $0_{a_{1}b_{1}}0_{b_{1}c_{1}}\cdots 0_{m_{1}n_{1}}$, $0_{a_{1}b_{1}}0_{b_{1}c_{1}}\cdots 1_{m_{1}n_{1}}$, $\cdots$, $1_{a_{1}b_{1}}1_{b_{1}c_{1}}\cdots 1_{m_{1}n_{1}}$. Therefore, the PCM results in the left side combined with the right side must be $0_{a_{1}b_{1}}0_{b_{1}c_{1}}\cdots 0_{m_{1}n_{1}}0_{a_{2}b_{2}}0_{b_{2}c_{2}}\cdots 0_{m_{2}n_{2}}$, $0_{a_{1}b_{1}}0_{b_{1}c_{1}}\cdots 1_{m_{1}n_{1}}0_{a_{2}b_{2}}0_{b_{2}c_{2}}\cdots 1_{m_{2}n_{2}}$, $\cdots$, or  $1_{a_{1}b_{1}}1_{b_{1}c_{1}}\cdots 1_{m_{1}n_{1}}1_{a_{2}b_{2}}1_{b_{2}c_{2}}\cdots 1_{m_{2}n_{2}}$.
 For example, if the PCM results is $0_{a_{1}b_{1}}0_{b_{1}c_{1}}\cdots 0_{m_{1}n_{1}}0_{a_{2}b_{2}}0_{b_{2}c_{2}}\cdots 0_{m_{2}n_{2}}$, the states
 $|\Phi^{\pm}_{1}\rangle_{N,2}$ will collapse to
 \begin{eqnarray}
&&|\Phi^{\pm}_{1}\rangle_{N,2} \rightarrow\frac{1}{2}[(|H\rangle_{a_{1}}|H\rangle_{a_{2}}|H\rangle_{b_{1}}|H\rangle_{b_{2}}\cdots|H\rangle_{n_{1}}|H\rangle_{n_{2}}\nonumber\\
&+&|V\rangle_{a_{1}}|V\rangle_{a_{2}}|V\rangle_{b_{1}}|V\rangle_{b_{2}}\cdots|V\rangle_{n_{1}}|V\rangle_{n_{2}})\nonumber\\
&\pm&(|H\rangle_{a_{1}}|V\rangle_{a_{2}}|H\rangle_{b_{1}}|V\rangle_{b_{2}}\cdots|H\rangle_{n_{1}}|V\rangle_{n_{2}}\nonumber\\
&+&|V\rangle_{a_{1}}|H\rangle_{a_{2}}|V\rangle_{b_{1}}|H\rangle_{b_{2}}\cdots|V\rangle_{n_{1}}|H\rangle_{n_{2}})]\nonumber\\
&=&\frac{1}{\sqrt{2}}(|H\rangle_{a_{1}}|H\rangle_{b_{1}}\cdots|H\rangle_{n_{1}}\pm|V\rangle_{a_{1}}|V\rangle_{b_{1}}\cdots|V\rangle_{n_{1}})\nonumber\\
&\otimes&\frac{1}{\sqrt{2}}(|H\rangle_{a_{2}}|H\rangle_{b_{2}}\cdots|H\rangle_{n_{2}}\pm|V\rangle_{a_{2}}|V\rangle_{b_{2}}\cdots|V\rangle_{n_{2}}).\nonumber\\
 \end{eqnarray}
 The second step is to distinguish $\frac{1}{\sqrt{2}}(|H\rangle_{a_{1}}|H\rangle_{b_{1}}\cdots|H\rangle_{n_{1}}+|V\rangle_{a_{1}}|V\rangle_{b_{1}}\cdots|V\rangle_{n_{1}})$ from
  $\frac{1}{\sqrt{2}}(|H\rangle_{a_{1}}|H\rangle_{b_{1}}\cdots|H\rangle_{n_{1}}-|V\rangle_{a_{1}}|V\rangle_{b_{1}}\cdots|V\rangle_{n_{1}})$.
 As shown in Fig. 3, after performing the Hadamard operation on each photons, we can make
  \begin{eqnarray}
 && \frac{1}{\sqrt{2}}(|H\rangle_{a_{1}}|H\rangle_{b_{1}}\cdots|H\rangle_{n_{1}}+|V\rangle_{a_{1}}|V\rangle_{b_{1}}\cdots|V\rangle_{n_{1}})\nonumber\\
  &\rightarrow&(\frac{1}{\sqrt{2}})^{N+1}[(|H\rangle_{a_{1}}+|V\rangle_{a_{1}})(|H\rangle_{b_{1}}+|V\rangle_{b_{1}})\nonumber\\
  &&\cdots(|H\rangle_{n_{1}}+|V\rangle_{n_{1}})\nonumber\\
  &&+(|H\rangle_{a_{1}}-|V\rangle_{a_{1}})(|H\rangle_{b_{1}}-|V\rangle_{b_{1}})\cdots(|H\rangle_{n_{1}}-|V\rangle_{n_{1}})],\nonumber\\
  \end{eqnarray}
  and
    \begin{eqnarray}
 && \frac{1}{\sqrt{2}}(|H\rangle_{a_{1}}|H\rangle_{b_{1}}\cdots|H\rangle_{n_{1}}-|V\rangle_{a_{1}}|V\rangle_{b_{1}}\cdots|V\rangle_{n_{1}})\nonumber\\
  &\rightarrow&(\frac{1}{\sqrt{2}})^{N+1}[(|H\rangle_{a_{1}}+|V\rangle_{a_{1}})(|H\rangle_{b_{1}}+|V\rangle_{b_{1}})\nonumber\\
  &&\cdots(|H\rangle_{n_{1}}+|V\rangle_{n_{1}})\nonumber\\
  &&-(|H\rangle_{a_{1}}-|V\rangle_{a_{1}})(|H\rangle_{b_{1}}-|V\rangle_{b_{1}})\cdots(|H\rangle_{n_{1}}-|V\rangle_{n_{1}})].\nonumber\\
  \end{eqnarray}
  After passing through the PBSs, if the number of $|V\rangle$ is even, it must be $ \frac{1}{\sqrt{2}}(|H\rangle_{a_{1}}|H\rangle_{b_{1}}\cdots|H\rangle_{n_{1}}+|V\rangle_{a_{1}}|V\rangle_{b_{1}}\cdots|V\rangle_{n_{1}})$, and the initial state is $|\Phi^{+}_{1}\rangle_{N,2}$. If the number of
  $|V\rangle$ is odd, it must be $\frac{1}{\sqrt{2}}(|H\rangle_{a_{1}}|H\rangle_{b_{1}}\cdots|H\rangle_{n_{1}}-|V\rangle_{a_{1}}|V\rangle_{b_{1}}\cdots|V\rangle_{n_{1}})$, and the initial state is  $|\Phi^{-}_{1}\rangle_{N,2}$.
 Certainly, if the PCM result is $1_{a_{1}b_{1}}0_{b_{1}c_{1}}\cdots 0_{m_{1}n_{1}}1_{a_{2}b_{2}}0_{b_{2}c_{2}}\cdots 0_{m_{2}n_{2}}$ in first step. The states
 $|\Phi^{\pm}_{1}\rangle_{N,2}$ will collapse to
  \begin{eqnarray}
&&|\Phi^{\pm}_{1}\rangle_{N,2}\nonumber\\
&\rightarrow&\frac{1}{\sqrt{2}}(|V\rangle_{a_{1}}|H\rangle_{b_{1}}\cdots|H\rangle_{n_{1}}\pm|H\rangle_{a_{1}}|V\rangle_{b_{1}}\cdots|V\rangle_{n_{1}})\nonumber\\
&\otimes&\frac{1}{\sqrt{2}}(|V\rangle_{a_{2}}|H\rangle_{b_{2}}\cdots|H\rangle_{n_{2}}\pm|H\rangle_{a_{2}}|V\rangle_{b_{2}}\cdots|V\rangle_{n_{2}}).\nonumber\\
 \end{eqnarray}
 It can be distinguished with the same method described above after performing a bit-flip operation on the photon in spatial mode $a_{1}$.
Such $2^{N-1}$ cases can be distinguished with the same principle.

Interestingly, if the initial states are $|\Phi^{\pm}_{2}\rangle_{N,2}$, the PCM result in $a_{1}b_{1}$ is always different from the PCM result
in  $a_{2}b_{2}$. The other PCM results in the left side equal to that in the right side. The PCM results in the left side have $2^{N-1}$ possible cases.
 The total PCM results can be written as
 $0_{a_{1}b_{1}}0_{b_{1}c_{1}}\cdots 0_{m_{1}n_{1}}1_{a_{2}b_{2}}0_{b_{2}c_{2}}\cdots 0_{m_{2}n_{2}}$,
  $0_{a_{1}b_{1}}0_{b_{1}c_{1}}\cdots 1_{m_{1}n_{1}}1_{a_{2}b_{2}}0_{b_{2}c_{2}}\cdots 1_{m_{2}n_{2}}$, $\cdots$, or
  $1_{a_{1}b_{1}}1_{b_{1}c_{1}}\cdots 1_{m_{1}n_{1}}0_{a_{2}b_{2}}1_{b_{2}c_{2}}\cdots 1_{m_{2}n_{2}}$.
 If the PCM results are   $0_{a_{1}b_{1}}0_{b_{1}c_{1}}\cdots 0_{m_{1}n_{1}}1_{a_{2}b_{2}}0_{b_{2}c_{2}}\cdots 0_{m_{2}n_{2}}$, the states
 $|\Phi^{\pm}_{2}\rangle_{N,2}$ will project to
   \begin{eqnarray}
  && |\Phi^{\pm}_{2}\rangle_{N,2}\nonumber\\
 &\rightarrow&\frac{1}{\sqrt{2}}(|H\rangle_{a_{1}}|H\rangle_{b_{1}}\cdots|H\rangle_{n_{1}}\pm|V\rangle_{a_{1}}|V\rangle_{b_{1}}\cdots|V\rangle_{n_{1}})\nonumber\\
&\otimes&\frac{1}{\sqrt{2}}(|V\rangle_{a_{2}}|H\rangle_{b_{2}}\cdots|H\rangle_{n_{2}}\pm|H\rangle_{a_{2}}|V\rangle_{b_{2}}\cdots|V\rangle_{n_{2}}).\nonumber\\
 \end{eqnarray}
 The second step is also to distinguish the states
 $\frac{1}{\sqrt{2}}(|H\rangle_{a_{1}}|H\rangle_{b_{1}}\cdots|H\rangle_{n_{1}}\pm|V\rangle_{a_{1}}|V\rangle_{b_{1}}\cdots|V\rangle_{n_{1}})$,
 which has been described in the former.

 If the initial states are $|\Phi^{\pm}_{3}\rangle_{N,2}$, we can find that both the PCM results in $a_{1}b_{1}$ and $b_{1}c_{1}$ are different from
 $a_{2}b_{2}$ and $b_{2}c_{2}$, respectively, while the other PCM results in the left side equal to that in the right side.
 For example, if the PCM result is $0_{a_{1}b_{1}}0_{b_{1}c_{1}}0_{c_{1}d_{1}}\cdots 0_{m_{1}n_{1}}1_{a_{2}b_{2}}1_{b_{2}c_{2}}0_{c_{2}d_{2}}\cdots 0_{m_{2}n_{2}}$,
 the states $|\Phi^{\pm}_{3}\rangle_{N,2}$ will project to
  \begin{eqnarray}
  && |\Phi^{\pm}_{3}\rangle_{N,2}\nonumber\\
 &\rightarrow&\frac{1}{\sqrt{2}}(|H\rangle_{a_{1}}|H\rangle_{b_{1}}\cdots|H\rangle_{n_{1}}\pm|V\rangle_{a_{1}}|V\rangle_{b_{1}}\cdots|V\rangle_{n_{1}})\nonumber\\
&\otimes&\frac{1}{\sqrt{2}}(|H\rangle_{a_{2}}|V\rangle_{b_{2}}|H\rangle_{c_{2}}\cdots|H\rangle_{n_{2}}\nonumber\\
&\pm&|V\rangle_{a_{2}}|H\rangle_{b_{2}}|V\rangle_{c_{2}}\cdots|V\rangle_{n_{2}}).
 \end{eqnarray}
 The next step is also to distinguish the states $\frac{1}{\sqrt{2}}(|H\rangle_{a_{1}}|H\rangle_{b_{1}}\cdots|H\rangle_{n_{1}}\pm|V\rangle_{a_{1}}|V\rangle_{b_{1}}\cdots|V\rangle_{n_{1}})$.
 If it is $\frac{1}{\sqrt{2}}(|H\rangle_{a_{1}}|H\rangle_{b_{1}}\cdots|H\rangle_{n_{1}}+|V\rangle_{a_{1}}|V\rangle_{b_{1}}\cdots|V\rangle_{n_{1}})$, the initial state must be $|\Phi^{+}_{3}\rangle_{N,2}$, otherwise, it must be $|\Phi^{+}_{3}\rangle_{N,2}$.

 If the initial is the arbitrary stats $|\Phi^{\pm}_{K}\rangle_{N,2}$ ($K=1, 2, \cdots N-1$), it can be distinguished in the same way. In this way,
 we can completely distinguish the C-GHZ state as shown in Eq. (\ref{multi3}).

 \section{discussion}
So far, we have fully described our logic Bell-state and C-GHZ state analysis. In the Bell state analysis, three PCM gates are required. In the first step,
two PCM operations on the $a_{1}b_{1}$ and $a_{2}b_{2}$ spatial modes are both performed. According to the measurement results, we can distinguish the states $|\Phi^{\pm}\rangle$ from $|\Psi^{\pm}\rangle$. If the measurement results are the same, the original states must be $|\Phi^{\pm}\rangle$. Otherwise, the original states must be
$|\Psi^{\pm}\rangle$. In the second step, we only need to distinguish the conventional polarized Bell state $|\phi^{+}\rangle$ from $|\phi^{-}\rangle$, which can
also be well distinguished with PCM gate in $a_{1}b_{1}$ modes. In this way, the four logic Bell states can be completely distinguished.
We showed that the arbitrary C-GHZ state can also be completely distinguished in the same way. As shown in Fig. 3, in the first step, both the left side
and right side perform $N-1$ PCM operations. According to the PCM operations, we can judge that the initial states must be one of the following states, say
$|\Phi^{\pm}_{1}\rangle_{N,2}$, $|\Phi^{\pm}_{2}\rangle_{N,2}$, $\cdots$, $|\Phi^{\pm}_{2^{N-1}}\rangle_{N,2}$. From Eq. (\ref{multi3}), if the logic qubit
is the same as  the neighbor one, the corresponded PCM results in the left side must be the same as that in the right side. Otherwise, the corresponded PCM
results must be different. For example,
the states $|\Phi^{\pm}_{1}\rangle_{N,2}$ will make all the PCM results be the same. However, if the states is $|\Phi^{\pm}_{2}\rangle_{N,2}$,
we can find that the first logic qubit in the spatial modes $a_{1}a_{2}$ is always different from the second logic qubit in the spatial modes $b_{1}b_{2}$.
Therefore, the PCM result
in the left side $a_{1}b_{1}$ must be different from it is in the right side  $a_{2}b_{2}$.
Certainly, we should point out that it has $2^{N-1}$ possible PCM results in all the left side in $a_{1}b_{1}$, $b_{1}c_{1}$, $\cdots$, $m_{1}n_{1}$.
 If the initial states are $|\Phi^{\pm}_{1}\rangle_{N,2}$, the PCM results in
$a_{1}b_{1}$ must be different from that in $a_{2}b_{2}$, while the other $2^{N-1}-1$ PCM results in the left side are the same as that in the right side.
In this way, in the first step, we can divide all the $2^{N}$ states into $2^{N-1}$ groups. In each group, there are two states,
 such as $|\Phi^{\pm}_{1}\rangle_{N,2}$, $|\Phi^{\pm}_{2}\rangle_{N,2}$, etc, as shown in Eq. (\ref{multi3}). In the second step,
 we are only required to distinguish
the conventional polarized GHZ states
 $\frac{1}{\sqrt{2}}(|H\rangle_{a_{1}}|H\rangle_{b_{1}}\cdots|H\rangle_{n_{1}}\pm|V\rangle_{a_{1}}|V\rangle_{b_{1}}\cdots|V\rangle_{n_{1}})$.
 After performing the Hadamard operations, they can be completely distinguished according to the number of $|V\rangle$ state.
 If the number of $|V\rangle$ is even, it must be
$\frac{1}{\sqrt{2}}(|H\rangle_{a_{1}}|H\rangle_{b_{1}}\cdots|H\rangle_{n_{1}}+|V\rangle_{a_{1}}|V\rangle_{b_{1}}\cdots|V\rangle_{n_{1}})$.
Otherwise, it must be $\frac{1}{\sqrt{2}}(|H\rangle_{a_{1}}|H\rangle_{b_{1}}\cdots|H\rangle_{n_{1}}-|V\rangle_{a_{1}}|V\rangle_{b_{1}}\cdots|V\rangle_{n_{1}})$.
In this way, all the C-GHZ states can be completely distinguished.

In our protocol, we discussed the logic Bell-state and C-GHZ state analysis. The logic qubit is encoded in the polarized Bell states
say $|\phi^{+}\rangle$ and $|\phi^{-}\rangle$. Actually.  the logic qubit in generalized concatenated GHZ state can be the $M$-particle GHZ state.
Therefore, the generalized logic Bell states  can be written as
\begin{eqnarray}
|\Phi_{M}^{\pm}\rangle_{AB}&=&\frac{1}{\sqrt{2}}(|GHZ_{M}^{+}\rangle_{A}|GHZ_{M}^{+}\rangle_{B}\nonumber\\
&\pm&|GHZ_{M}^{-}\rangle_{A}|GHZ_{M}^{-}\rangle_{B}),\nonumber\\
|\Psi_{M}^{\pm}\rangle_{AB}&=&\frac{1}{\sqrt{2}}(|GHZ_{M}^{+}\rangle_{A}|GHZ_{M}^{-}\rangle_{B}\nonumber\\
&\pm&|GHZ_{M}^{-}\rangle_{A}|GHZ_{M}^{+}\rangle_{B}).\label{logicbell}
\end{eqnarray}
The generalized C-GHZ state can be written as
\begin{eqnarray}
|\Phi^{\pm}_{1}\rangle_{N,M}&=&\frac{1}{\sqrt{2}}(|GHZ_{M}^{+}\rangle^{\otimes N}\pm|GHZ_{M}^{-}\rangle^{\otimes N}),\nonumber\\
|\Phi^{\pm}_{2}\rangle_{N,M}&=&\frac{1}{\sqrt{2}}(|GHZ_{M}^{-}\rangle|GHZ_{M}^{+}\rangle^{\otimes N-1}\nonumber\\
&\pm&|GHZ_{M}^{+}\rangle|GHZ_{M}^{-}\rangle^{\otimes N-1}),\nonumber\\
&\cdots&\nonumber\\
|\Phi^{\pm}_{2^{N-1}}\rangle_{N,M}&=&\frac{1}{\sqrt{2}}(|GHZ_{M}^{+}\rangle^{\otimes N-1}|GHZ_{M}^{-}\rangle\nonumber\\
&\pm&|GHZ_{M}^{-}\rangle^{\otimes N-1}|GHZ_{M}^{+}\rangle.\label{multi4}
\end{eqnarray}

Interestingly, the logic Bell states and the GHZ states shown in Eqs. (\ref{logicbell}) and (\ref{multi4}) can also be completely distinguished with the same method.
The logic Bell state analysis described in  Eq. (\ref{logicbell}) equals to that described in Eq.(\ref{bell}) and the logic GHZ
state analysis described in Eq. (\ref{multi4}) also equals to that described in Eq. (\ref{multi2}). Due to the state in Eq. (\ref{logicbell})
and the state in Eq.(\ref{bell}), the state in Eq. (\ref{multi4}) and the state in Eq. (\ref{multi2}) have the same logic structure, respectively.  By measuring the photons in number 3 to $M$ in the basis
$|\pm\rangle=\frac{1}{\sqrt{2}}(|H\rangle\pm|V\rangle)$ in each logic qubit, the states $|\Phi_{M}^{\pm}\rangle_{AB}$ will become
$|\Phi^{\pm}\rangle_{AB}$, and $|\Psi_{M}^{\pm}\rangle_{AB}$ will become
$|\Psi^{\pm}\rangle_{AB}$ if the number of $|-\rangle$ is even. Otherwise, $|\Phi_{M}^{\pm}\rangle_{AB}$ will become
$|\Phi^{\mp}\rangle_{AB}$, and $|\Psi_{M}^{\pm}\rangle_{AB}$ will become
$|\Psi^{\mp}\rangle_{AB}$. In the next step, we should perform the logic Bell state analysis described in Sec.III. For the C-GHZ state analysis
described in Eq. (\ref{multi4}), we can use the same approach to simplify them to the states shown in Eq. (\ref{multi2}). In this way, the arbitrary
C-GHZ state can be completely distinguished.

In our protocol, the key element to  realize the Bell-state analysis is the PCM gate, which constructed by the cross-Kerr nonlinearity. Though there are
many  theoretical works for quantum information processing based on cross-Kerr nonlinearity,
the cross-Kerr nonlinearity is still a controversial topic \cite{Gea,Shapiro1,Shapiro2,kok1,kok2}. The debate over the usefulness of photonic quantum
information processing based on the cross-Kerr nonlinearity is that the phase shift is too small to be measured  in a single photon level.
Recently, some researches showed that it is possible to obtain the observable value of the Kerr phase shift \cite{hofmann,weak_meaurement,oe,he2}.

\section{conclusion}
In conclusion, we have described a two-step  approach to realize the complete logic Bell-state and arbitrary C-GHZ state analysis. In our protocol, we exploit the cross-Kerr nonlinearity
to construct the PCM gate. With the help of PCM gates, the whole task can be divided into two steps. In the first step, after performing the PCM
operations, the four states can be divided in two groups. The first group is $\{|\Phi^{\pm}\rangle\}$ and the second group is $\{|\Psi^{\pm}\rangle\}$.
In the second step, the states $|\Phi^{\pm}\rangle$ and $|\Psi^{\pm}\rangle$ in each group can also be discriminated by PCM operation. Our protocol can  be
extended to distinguish the arbitrary C-GHZ state. It can also be divided into two steps. In the first step, all the $2^{N}$ C-GHZ states can be divided into $2^{N-1}$
groups, according to the different PCM results in both left and right side. In each group, the two states can also be
completely distinguished in the second step. Our protocol has its practical application for future long-distance quantum communication based on logic
qubit entanglement.

\section*{ACKNOWLEDGEMENTS}
This work is supported by the National Natural Science Foundation of China (Grant Nos. 11104159 and 11347110),
Qing Lan Project, Jiangsu Province, 1311 Talent Plan, Nanjing University of Posts and Telecommunications,
 and the Priority Academic Development Program of Jiangsu Higher Education Institutions, China.

\end{document}